\documentclass[preprint]{emulateapj}
\begin{document}
\title{
Dissipative Pulsar Magnetosphere
}
\author{Andrei Gruzinov}
\affil{CCPP, Physics, NYU, 4 Washington Pl., New York, NY 10003}
\subjectheadings{pulsars -- neutron stars -- magnetospheres }
\begin{abstract}

Dissipative axisymmetric pulsar magnetosphere is calculated by a direct numerical simulation of the Strong-Field Electrodynamics equations. The magnetic separatrix disappears, it is replaced by a region of enhanced dissipation. With a better numerical scheme, one should be able to calculate the bolometric lightcurves for a given conductivity. 

\end{abstract}
\maketitle
\section{Introduction}
Recent progress in calculating pulsar magnetospheres (Spitkovsky 2006 and references therein) is based on Force-Free Electrodynamics (FFE). FFE describes electromagnetic fields of special geometry -- at each event, the electric field is smaller than and perpendicular to the magnetic field. When applied to pulsars, FFE predicts a singular current layer along the magnetic separatrix surface. Within the light cylinder, the singularity is of an admissible type -- although the current is singular, the electric field remains smaller than and perpendicular to the magnetic field. Outside the light cylinder, the singularity actually violates the FFE equations. For this reason, and also to understand dissipation processes leading to the observed radiation, one wants to use a dissipative generalization of FFE. Such a theory -- Strong-Field Electrodynamics (SFE) -- has recently been proposed (Gruzinov 2008).

\begin{figure}
\epsscale{1.25}
\plotone{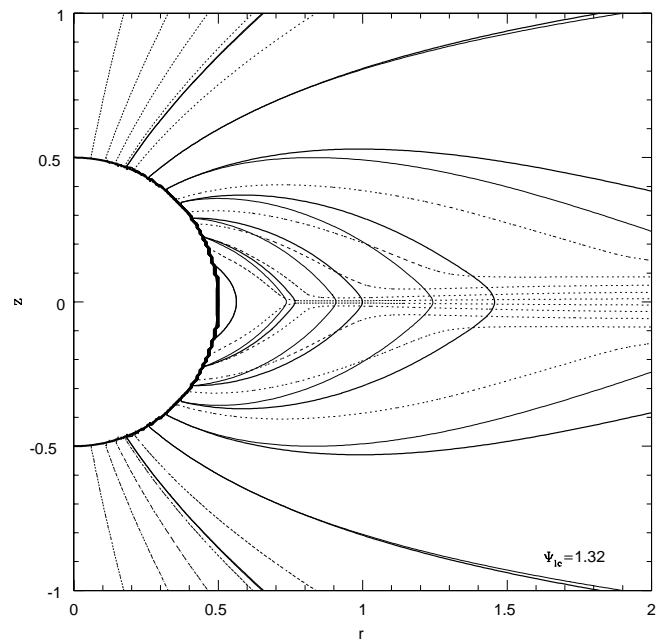}
\caption{$\sigma =10$, $r_s=0.5$.  Thick: magnetic stream function $\psi$, intervals of $\psi (r=1, \theta =\pi/2)/5$. Thin: electric potential $\phi$, same intervals. Dotted: poloidal current $I$, see Fig. 3 for current distribution on the surface of the star.}
\end{figure}

\begin{figure}
\epsscale{1.15}
\plottwo{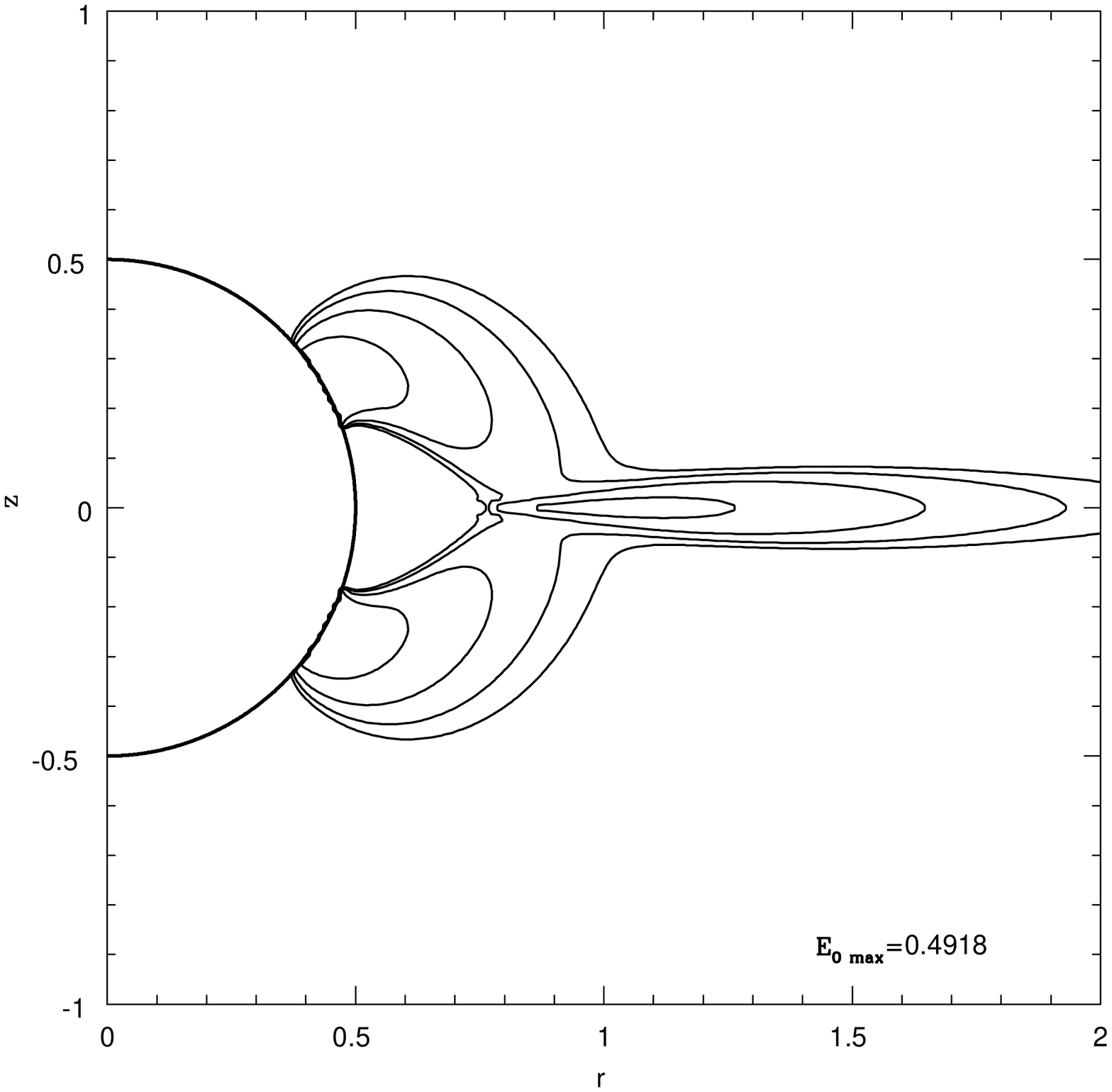}{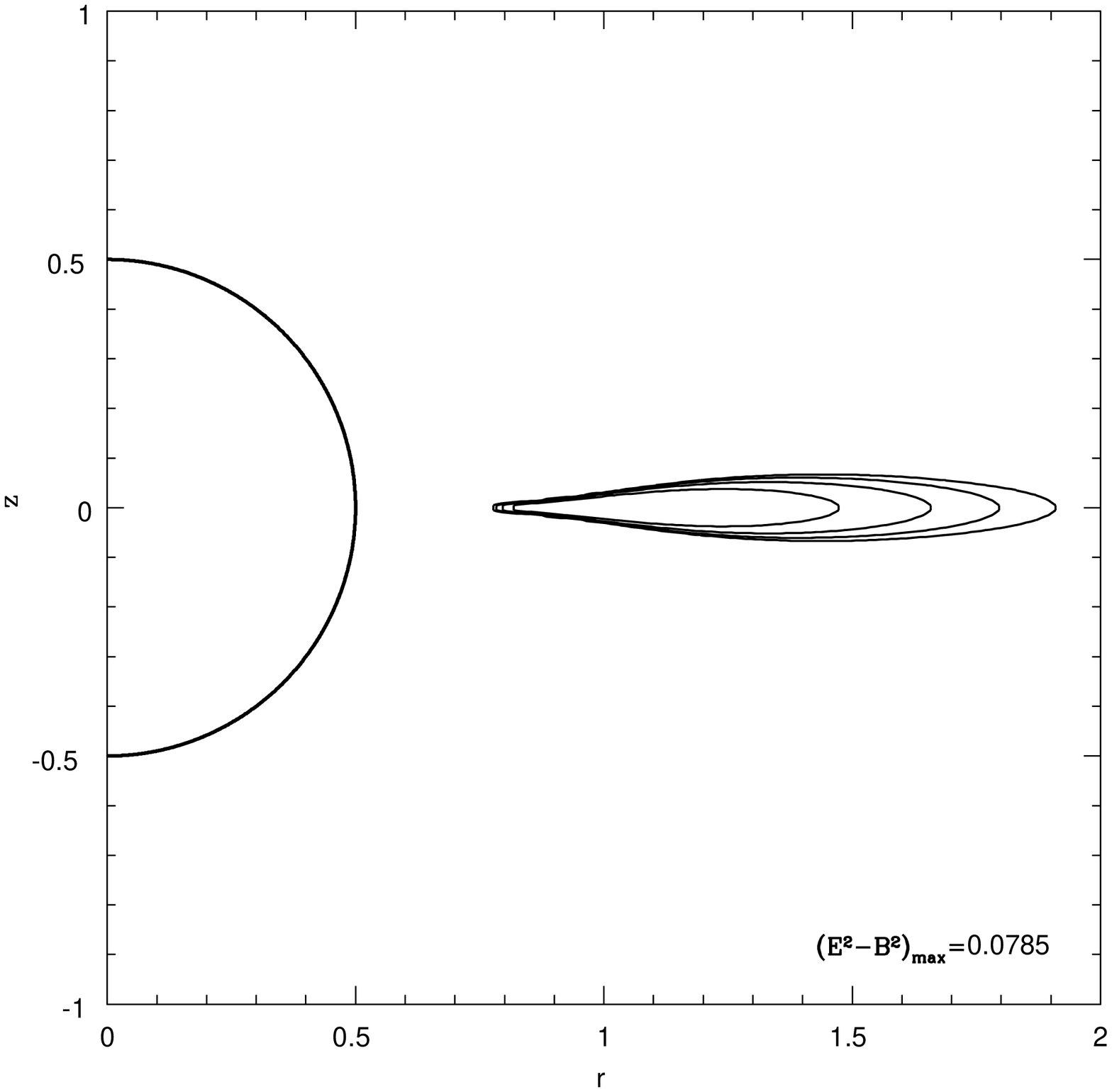}
\caption{{\it Left:} Invariant $E_0$, intervals $E_{0~max}/5$. {\it Right:} Invariant $E^2-B^2$, intervals $(E^2-B^2)_{max}/5$, showing only positive isolines. }
\end{figure}

\begin{figure}
\epsscale{1.15}
\plottwo{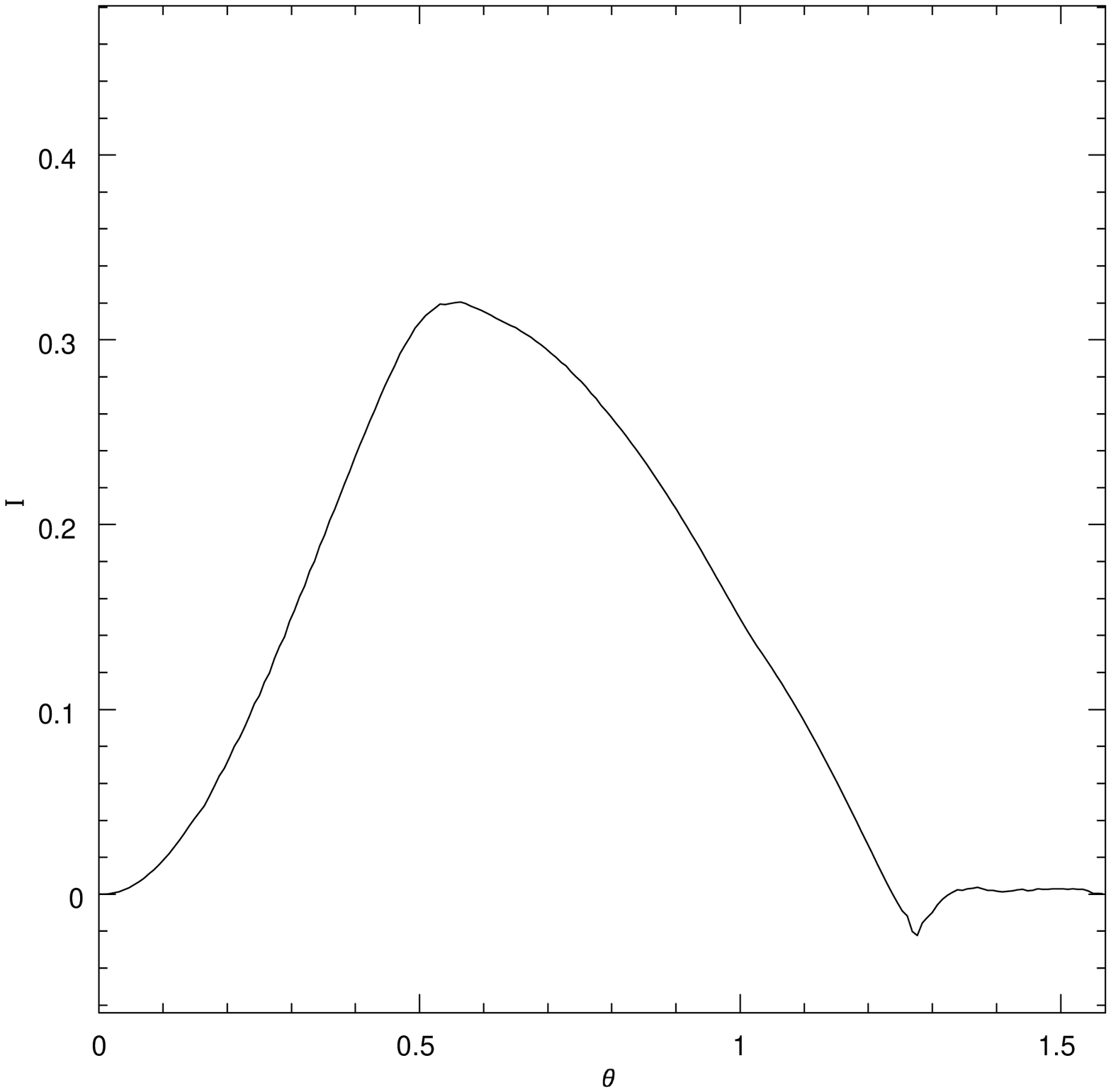}{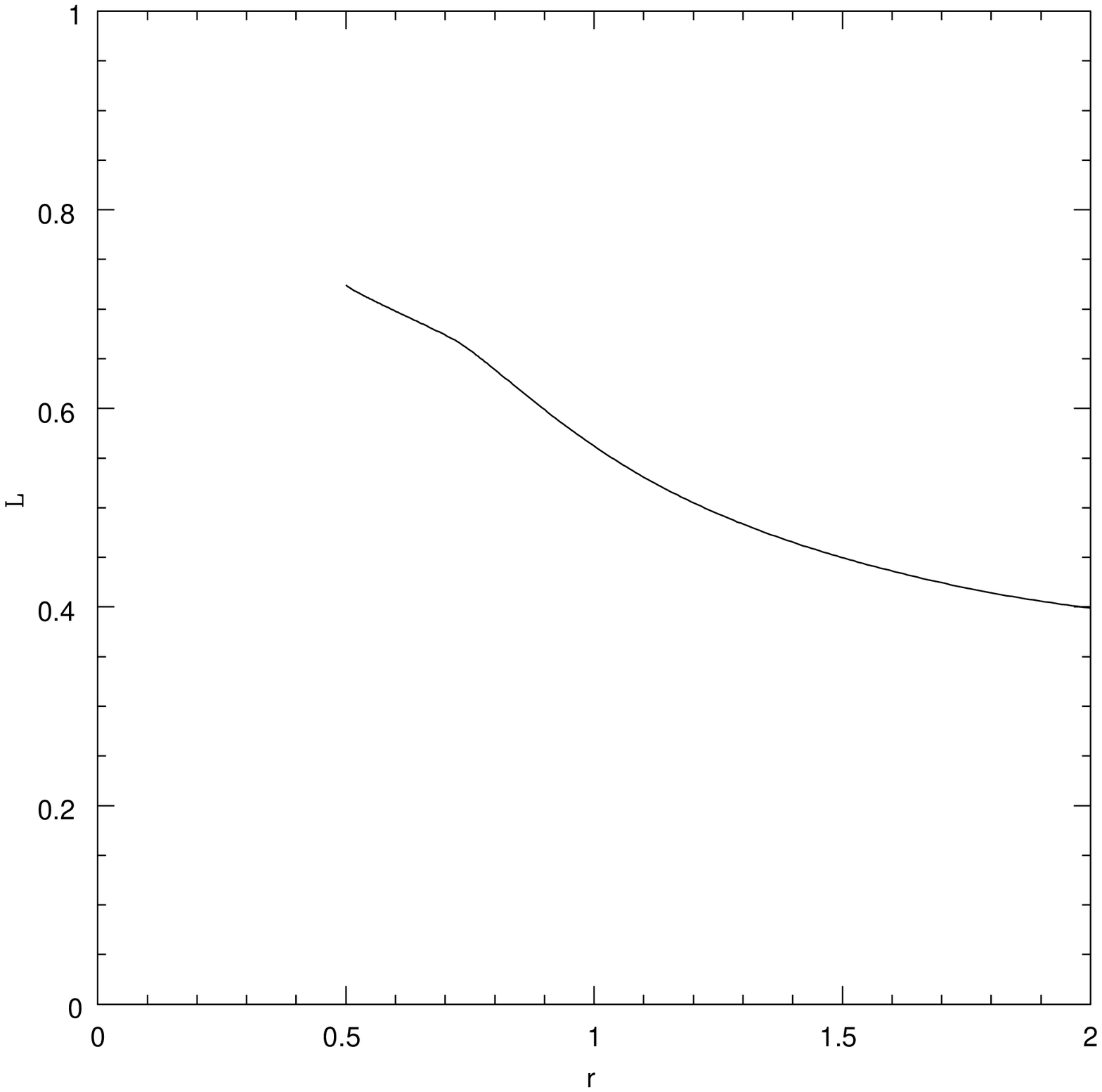}
\caption{{\it Left:} Poloidal current distribution over the surface of the star. {\it Right:} Poynting luminosity at different radii.}
\end{figure}

\begin{figure}
\epsscale{1.25}
\plotone{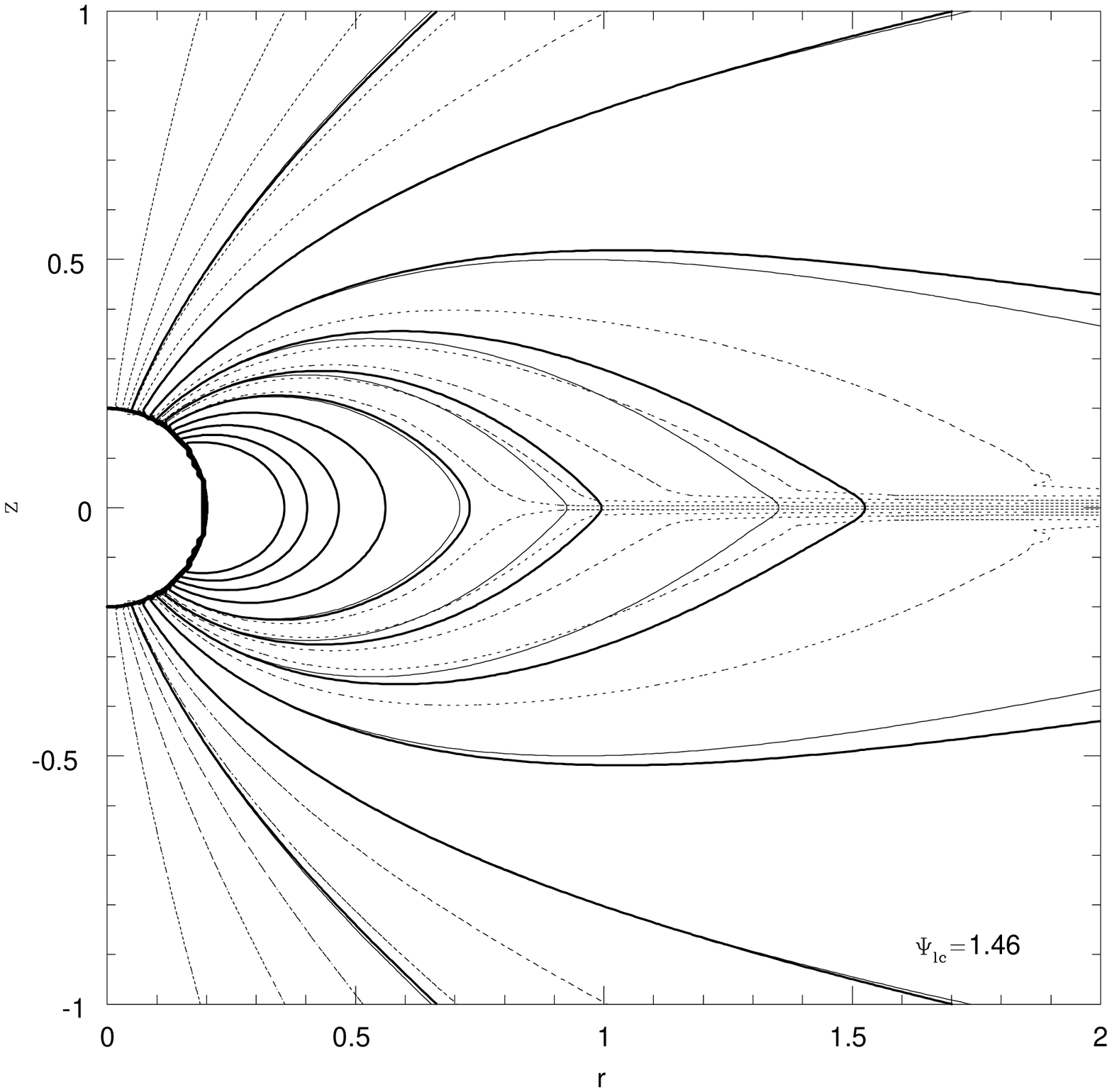}
\caption{Same as fig.1, $\sigma =50$, $r_s=0.2$. }
\end{figure}

\begin{figure}
\epsscale{1.15}
\plottwo{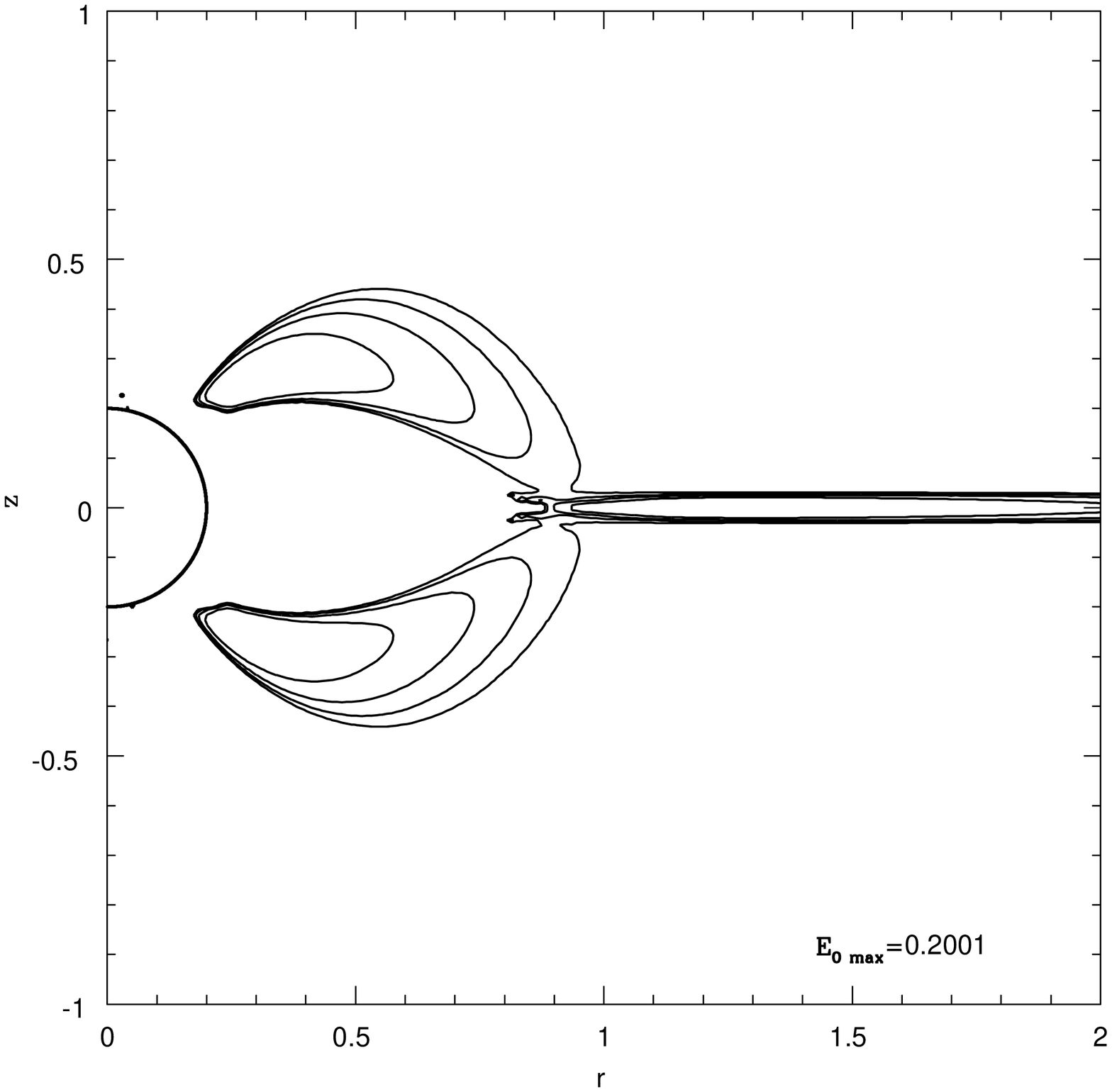}{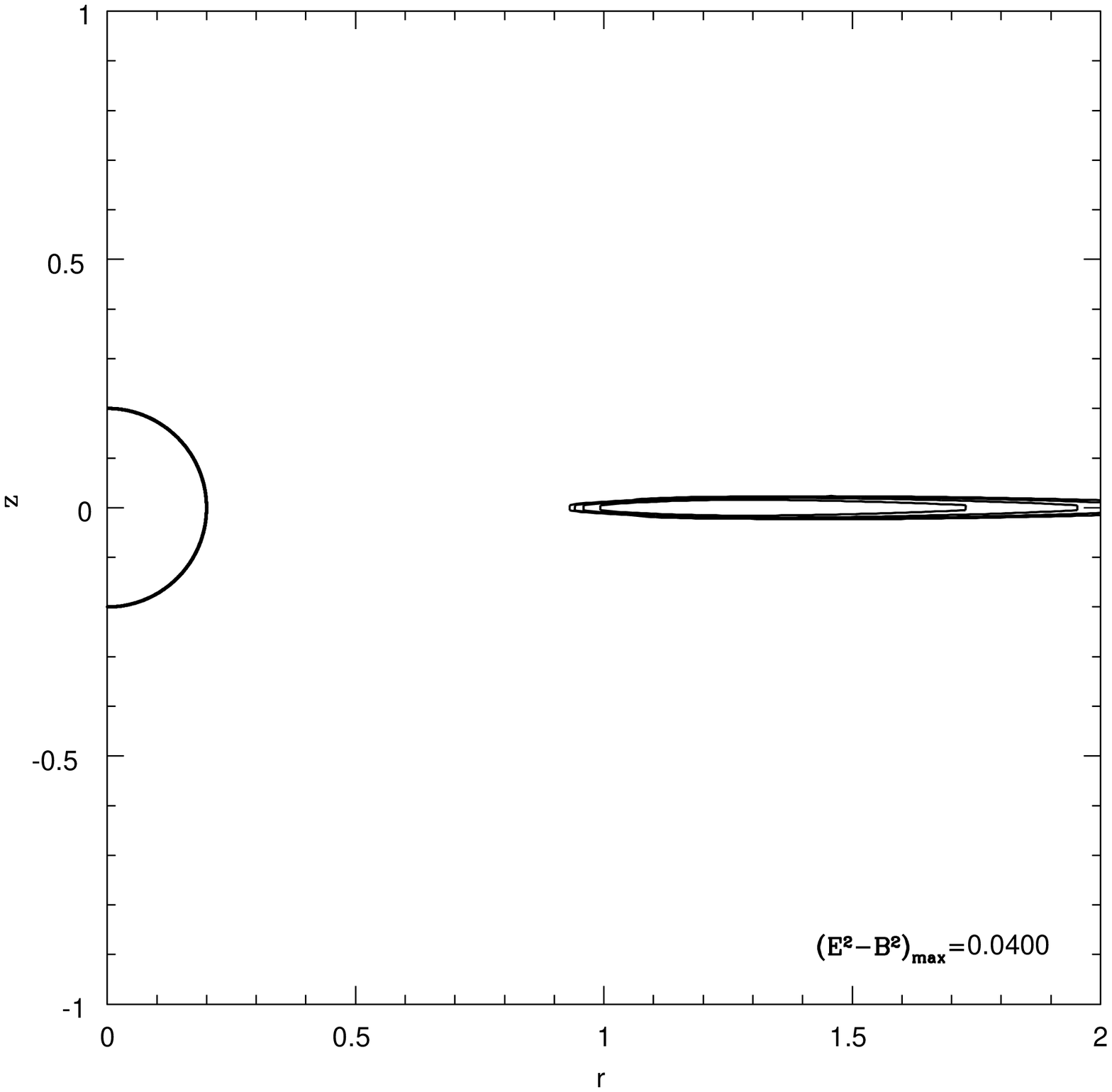}
\caption{Same as fig.2, $\sigma =50$, $r_s=0.2$. }
\end{figure}

SFE describes electromagnetic fields of arbitrary  geometry. SFE is Maxwell theory with the semi-dissipative Lorentz-covariant Ohm's law
\begin{equation}\label{ohmsfe}
{\bf j}={\rho {\bf E}\times {\bf B} +(\rho ^2+\gamma ^2\sigma ^2E_0^2)^{1/2}(B_0{\bf B}+E_0{\bf E})\over B^2+E_0^2},
\end{equation}
where $\rho \equiv \nabla \cdot {\bf E}$ is the charge density, $B_0^2-E_0^2\equiv {\bf B}^2-{\bf E}^2$, $B_0E_0\equiv {\bf E}\cdot {\bf B}$, $E_0\geq 0$ represent the field invariants, and $\gamma ^2 \equiv {B^2+E_0^2\over B_0^2+E_0^2}$. The conductivity scalar $\sigma=\sigma(B_0,E_0)$ is an arbitrary function of the field invariants. SFE comes from the following simple postulate: at each event, in the frame where the electric field is parallel to the magnetic field and the charge density vanishes, the current $\sigma E$ flows along the fields. \footnote{SFE handles time-like 4-currents (Gruzinov 2008).}. Covariant formulation of SFE is $j^2=-\sigma ^2E_0^2$, $B_0Fj=E_0\tilde{F}j$ (compare to FFE: $E_0=0$, $Fj=0$).

\section{Axisymmetric Pulsar in SFE}

We simulated time-dependent axisymmetric Maxwell equations with the SFE Ohm's law. We used the regularization described in (Gruzinov 2008), and checked that the results are independent of the regularization to the specified accuracy.

The initial field was vacuum around the rotating magnetized neutron star. In spherical coordinates, in pulsar units ($\mu =\Omega =c=1$), the vacuum field is given by the magnetic stream function $\psi =\sin ^2\theta /r$ and the electrostatic potential $\phi =(1/3-\cos ^2\theta )r_s^2/r^3$, where $r_s$ is the radius of the star. We then evolved the three components of the electric field $(E_r,E_{\theta }, E_{\phi})$, the magnetic stream function $\psi$, and the toroidal magnetic field $B_{\phi}$. The boundary conditions on the surface of the star are $\psi =\sin ^2\theta /r_s$ and $E_{\theta }=-\sin (2\theta )/r_s^2$.

We show the magnetosphere at $t=6$, when the field configuration within $r=2$ (two light cylinders) saturates. The outer boundary was at $r=5$ -- out of causal contact with the $r<2$ zone. The saturated electric field is potential. The saturated toroidal field is given by the poloidal current $B_\phi \equiv 2I/(r\sin \theta )$. 

We show two cases. The $r_s=0.5$, $\sigma =10$ results (for the Poynting power) are accurate to about 2\% . Fig.1. shows the field lines, Fig.2. -- the field invariants, Fig.3. -- the integrated poloidal current as a function of the latitude and the Poynting luminosity (in pulsar units) as a function of radius. The luminosity decreases due to dissipation. Figg. 4-6 show the $r_s=0.2$, $\sigma =50$ results; these are accurate to about 10\% .

\begin{figure}
\epsscale{1.15}
\plottwo{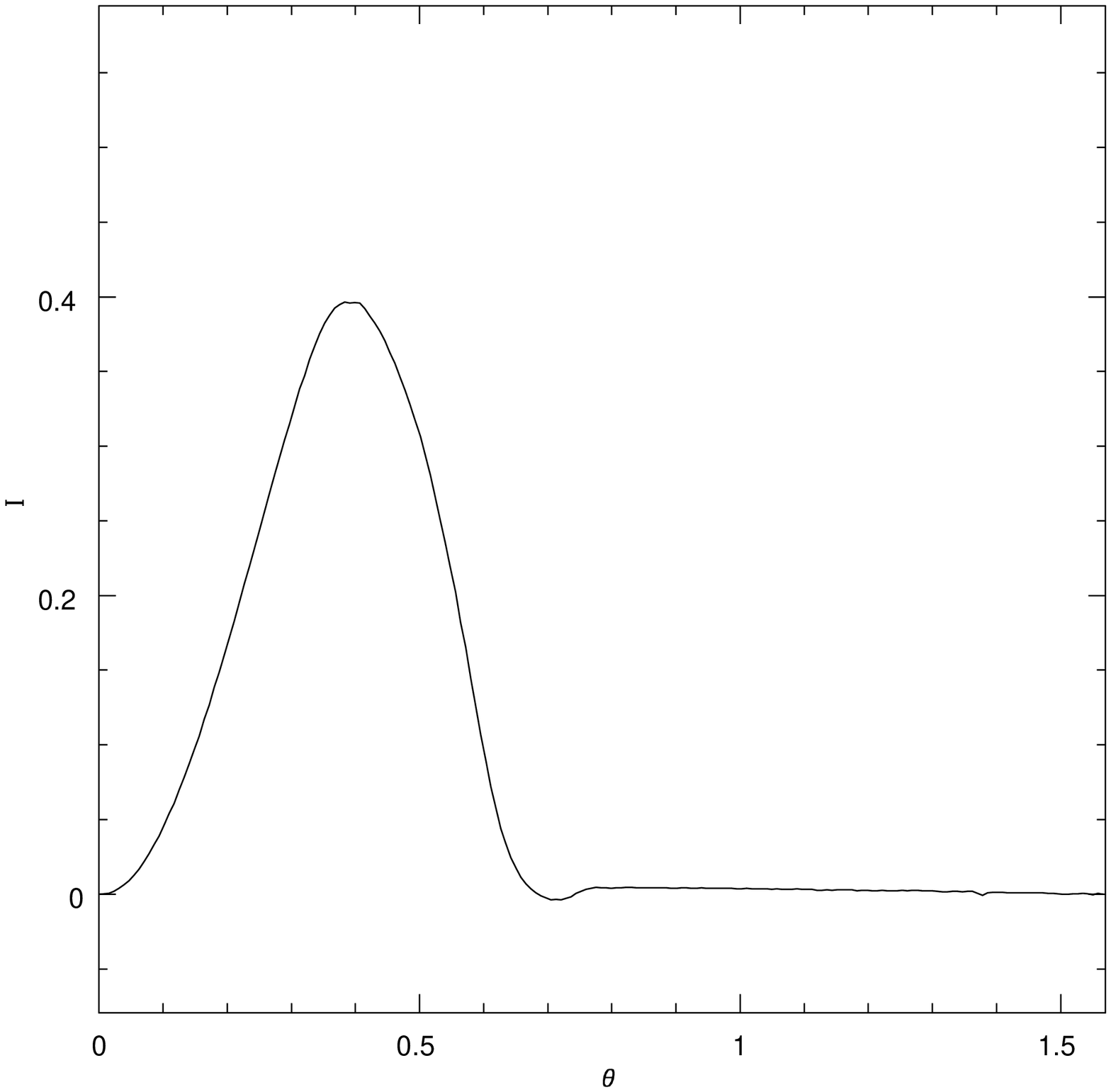}{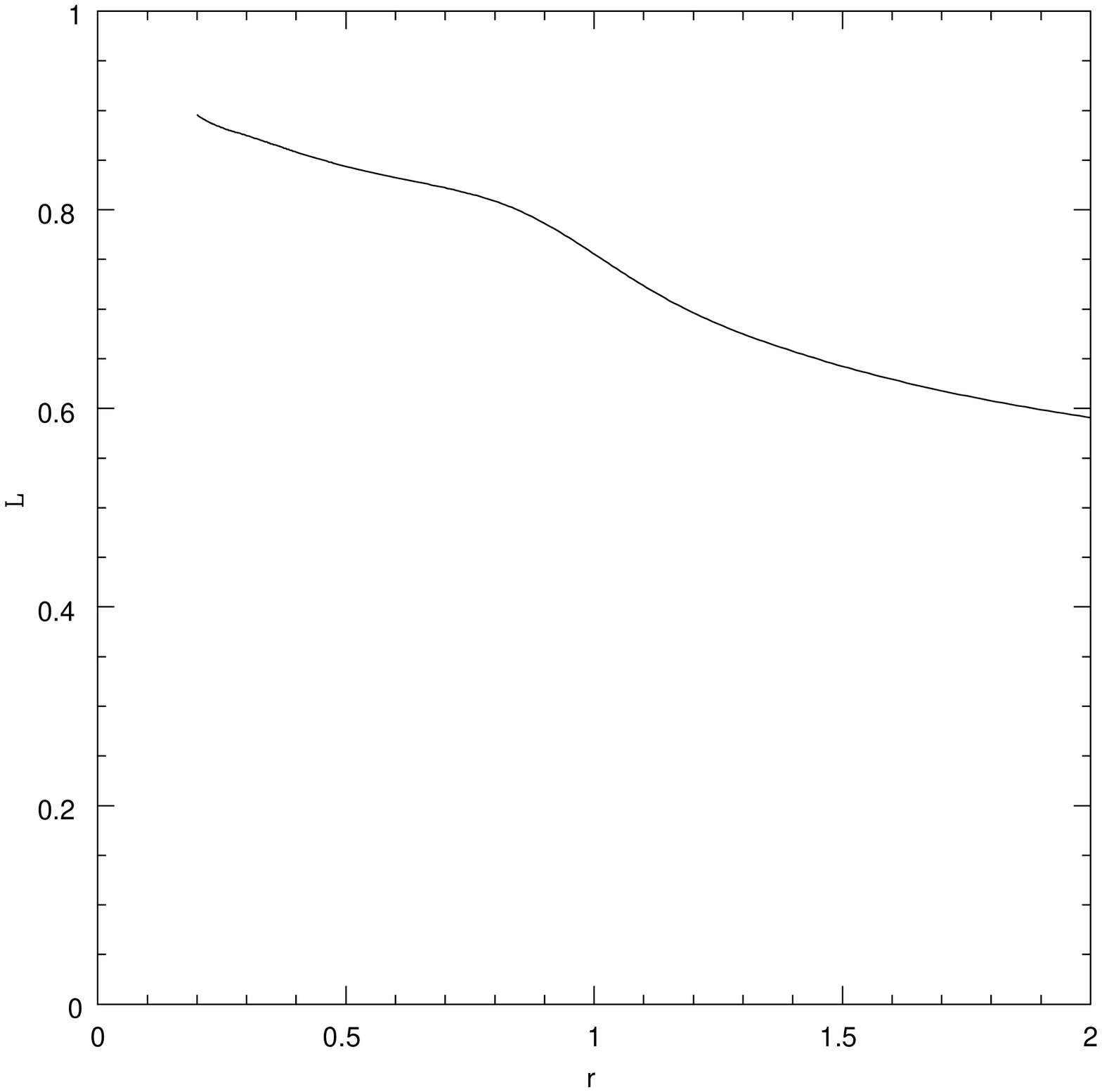}
\caption{Same as fig.3, $\sigma =50$, $r_s=0.2$.}
\end{figure}

\section{Conclusion}

Our code is amateurish  -- a straightforward discretization with diffusive stabilization \footnote{Available upon request from ag92@nyu.edu.}. A professional numerical scheme might be of interest for two reasons:

(i) Our pulsar, even at $\sigma =50$, is too dissipative. It would be interesting to increase $\sigma$ to the values when the return current becomes space-like. 

(ii) With a good 3D code, it should be possible to calculate the pulsar bolometric lightcurves for a given surface field and conductivity \footnote{ In SFE, $\sigma$ is an arbitrary function of the field invariants. In the real world, the star breaks Lorentz invariance. }. To calculate emission, one assumes efficient heat to light conversion and strong beaming along the field lines (in the frame where the fields are parallel).

\acknowledgements

I thank Andrew MacFadyen for useful advice. This work was supported by the David and Lucile Packard foundation.

\end{document}